\documentclass[amssymb,amsfonts,aps,prb,twocolumn]{revtex4}
\usepackage{graphicx}

\begin{document}

\title{Non-Collinearity in Small Magnetic Cobalt-Benzene Molecules}

\date{\today}

\author{J. W. Gonz\'alez$^{1,\,}$\footnote{Corresponding author: sgkgosaj@ehu.eus}, T. Alonso-Lanza$^{1}$, F. Delgado$^{1,\,2,\,3}$, F. Aguilera-Granja$^{1,4}$, A. Ayuela$^{1}$}
\affiliation{
$^{(1)}$ Centro de F\'{i}sica de Materiales (CSIC-UPV/EHU)-Material Physics Center (MPC), Donostia International Physics Center (DIPC), Departamento de F\'{i}sica de Materiales, Fac. Qu\'{i}micas UPV/EHU. Paseo Manuel de Lardizabal 5, 20018, San Sebasti\'an-Spain.
\\
$^{(2)}$ IKERBASQUE, Basque Foundation for Science, E-48013 Bilbao, Spain.
\\
$^{(3)}$ Departamento de F\'{i}sica, Universidad de La Laguna, La Laguna E-38204, Tenerife, Spain.
\\
$^{(4)}$ Instituto de F\'isica, Universidad Aut\'onoma de San Luis Potos\'i,  78000 San Luis Potos\'i, M\'exico.}

\begin{abstract}
Organometallic clusters based on transition metal atoms are interesting because possible 
applications in spintronics and quantum information. In addition to the enhanced magnetism 
at the nanoscale, the organic ligands may provide a natural shield again unwanted magnetic 
interactions with the matrices required for applications. Here we show that the organic 
ligands may lead to non-collinear magnetic order as well as the expected quenching of the 
magnetic moments.
We use different density functional theory (DFT) methods to study the experimentally 
relevant three cobalt atoms surrounded by benzene rings (Co$_3$Bz$_3$).
We found that the benzene rings induce a ground state with non-collinear magnetization, 
with the magnetic moments localized on the cobalt centers and lying on the plane formed 
by the three cobalt atoms. 
We further analyze the magnetism of such a cluster using an anisotropic Heisenberg model 
where the involved parameters are obtained by a comparison with the DFT results. These 
results may also explain the recent observation of null magnetic moment of Co$_3$Bz$_3^+$.
Moreover, we propose an additional experimental verification based on electron paramagnetic 
resonance.
\end{abstract}

\maketitle

\section{\label{sec:intro} Introduction}

\begin{figure}[hb!]
\centering
\includegraphics[clip=true,width=.3\textwidth]{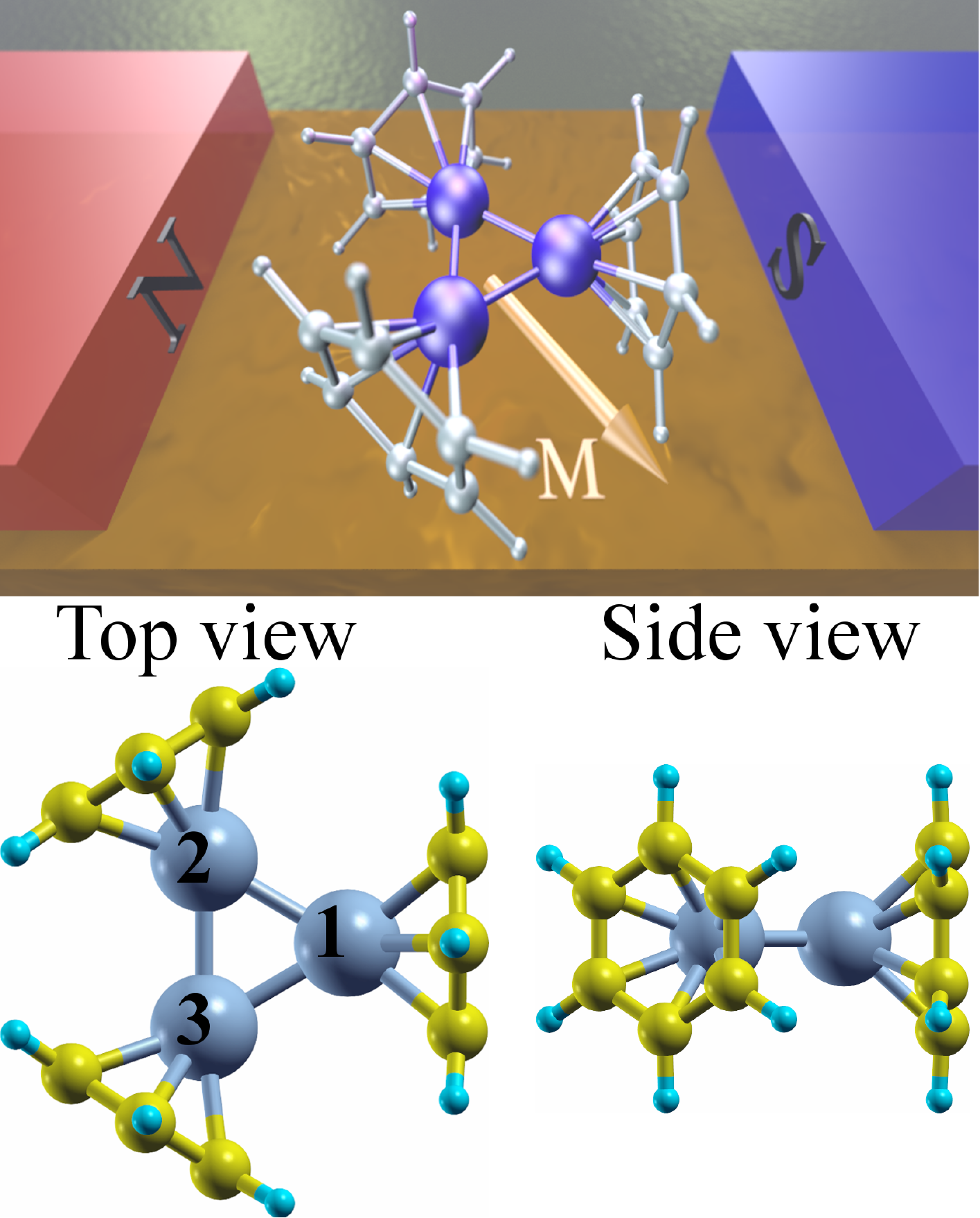}
\caption{(Color online) Upper panel, scheme of the Co$_3$Bz$_3$ cluster deposited on a substrate. 
Lower panel, most stable configuration of the cluster. The cobalt atoms (light blue)  form an 
equilateral triangle, and the benzene rings (carbon shown in yellow and hydrogen in cyan) are 
perpendicular to the plane formed by the cobalt atoms.}
\label{fig1:esq}
\end{figure}
 
In recent years, a great deal of attention has been devoted to  magnetic organometallic 
compounds  due to their prospective use as nanoscale magnetic devices. 
For example, single-molecule magnets (SMMs)\cite{bogani2008molecular,Molen_Sense_natnano_2013} 
are currently considered as potential building blocks for 
spintronics,\cite{Ardavan_Rival_prl_2007,Bertaina_Gambarelli_nature_2008} molecular 
electronics,\cite{Molen_Sense_natnano_2013} and quantum 
computing.\cite{leuenberger2001quantum,ardavan2009storing,mannini2009magnetic}
Magnetic  organometallic molecules can be formed by bringing together small clusters of 
transition metal atoms and benzene molecules, such as in the case of cobaltocene and ferrocene.%
Their magnetic moments have been measured using Stern-Gerlach molecular-beam deflections showing 
that part of the original magnetization may survive when the number of transition metal
centers and ligands are comparable. \cite{scully1987theory,knickelbein2004magnetic,payne2007magnetic} 
More recently, X-ray magnetic circular dichroism spectroscopy (XMCD) has been used to determine 
independently the orbital- and spin-magnetic moments measured on a trapped and cooled gas of metal cluster ions.\cite{zamudio2015direct,langenberg2014spin,meyer2015spin}
Contrary to the SMMs, where the transition metal centers are surrounded by organic compounds 
and thus, protecting their magnetic moment,\cite{candini2011graphene,layfield2014organometallic,galloway2008cobalt} 
the transition metal atoms in such organometallic clusters are in close proximity, leading to 
strong exchange interactions.
Although these organometallic clusters made up of transition metal clusters and benzene rings 
share same magnetic properties with SMMs, their magnetism is much less studied.

Cobalt-based clusters constitute prototypes for organometallic clusters since they present predominantly 
ferromagnetic order and typically have high magnetic anisotropies.
Experimental results suggest that clusters of late transition metals fully saturated with benzene 
molecules are stable $\pi$-complexes that are not very 
reactive.\cite{kurikawa1995structures,kurikawa1997structures,kurikawa1999electronic}
Rather than having the metal center(s) sandwiched between benzene rings, as was typical in 
early transition metal complexes,\cite{wedderburn2006geometries,kurikawa1999electronic,kurikawa1995structures,kurikawa1997structures}  
the cobalt centers in cobalt-benzene (Co-Bz) clusters are close together, forming an 
inner metallic cluster bound through $\pi$-bonds to benzene molecules around the cluster, 
commonly called rice-ball configuration.

The magnetic properties of Co-Bz complexes have been measured in Stern-Gerlach
 molecular beam deflection experiments between $60$ K and $310$ K.\cite{knickelbein2006magnetic,payne2007magnetic} 
The magnetic moments per cobalt atom of Co$_n$Bz$_m$  clusters 
with $2 \leq n\leq 200$, have been found to range from $3.5\, \mu_B $ to $1.8\, \mu_B $ 
and to decrease as the cluster size increases. 
These magnetic moments were all higher than the  bulk cobalt magnetic moment of $1.72\,\mu_B$. 
However, marked discrepancies between magnetic moments determined in different experiments have been reported.\cite{knickelbein2006magnetic,payne2007magnetic}
The magnetism of ionized Co$_3$Bz$_m^+$ ($m=0,1,2,3$) has recently been studied using XMCD at cryogenic conditions looking at orbital- and spin-magnetic moments. 
Surprisingly, Co$_3$Bz$_3^+$ displays not magnetic response at all.
By using collinear DFT calculations they tried to explain the origin of the quenched magnetic moment, but the results were highly sensible to the exchange correlation functionals used.\cite{akin2016size}
Henceforth, the magnetic properties of  Co-Bz  clusters  and  their  response  to  magnetic  fields require further research.

 Thus, the aim of this work is to theoretically examine these properties taking the meaningful example of  Co$_3$Bz$_3$ cluster.  
We consider Co$_3$Bz$_3$ due to its small size and its closed
 structure. It should be noted that  Co$_3$Bz$_3$ clusters have been found to be more stable and abundant than other sizes, a fact that can be analyzed by mass spectrometry.\cite{wedderburn2006geometries,kurikawa1999electronic,kurikawa1995structures,kurikawa1997structures} 
In our study we explore both the collinear and non-collinear magnetic orders. We  used several  density functional theory (DFT) approaches. For our most accurate and detailed description, 
we used a relativistic and full-potential linearised augmented-plane wave (FP-LAPW) \cite{singh2006planewaves,sjostedt2000alternative} including the effects of spin-orbit interactions.
Our results show that magnetic coupling between the cobalt atoms is strongly modified by the benzene molecules.
Crucially, we found that the minimum energy configuration corresponds to a non-collinear order in which the
magnetic moments lie in the plane defined by the three cobalt atoms, which is almost $200$ meV  below the first excited ferromagnetic configuration.
Finally, based on the results of the DFT calculations, we introduced a minimum quantum-spin model that accounts for both, magnetic anisotropies and inter-spin couplings. The results of the model suggest that the most stable configurations corresponds to a non-symmetric exchange interaction between the three cobalt atoms with a significant local anisotropy due to the benzene rings.
Thus, the resulting magnetic landscape emerges both, from the antiferromagnetic coupling between the CoBz clusters and the closed structure of the Co$_3$Bz$_3$ molecule, which lead to the existence of non-trivial magnetic orders and geometrical spin frustration.

\section{Results on Magnetic Order}
We  investigated  the relaxed structures using electronic structure calculations resolved for spin within the DFT framework. The Kohn-Sham ansatz uses a linear combination of atomic orbitals, as used in the SIESTA code.\cite{soler2002siesta}
The relaxed structures produced using SIESTA were successfully tested and refined using two other plane-wave codes, VASP\cite{VASP,VASP2} and  {\sc Quantum ESPRESSO}.\cite{QE-2009}  We started with different input geometries and determined how the magnetic properties changed. 
In general, describing the magnetic properties of systems including transition metals in detail requires a proper relativistic treatment of
electron-nuclear and electron-electron interactions.
We deal with spin-orbit interactions using an accurate relativistic version of the full-potential linearized augmented plane-wave method\cite{singh2006planewaves,sjostedt2000alternative} as it is implemented in the ELK-code.\cite{ELK}
A more detailed information on the DFT calculations is provided in the Supplemental Material.

\begin{figure}[hb!]
\includegraphics[clip,width=0.48\textwidth,angle=0]{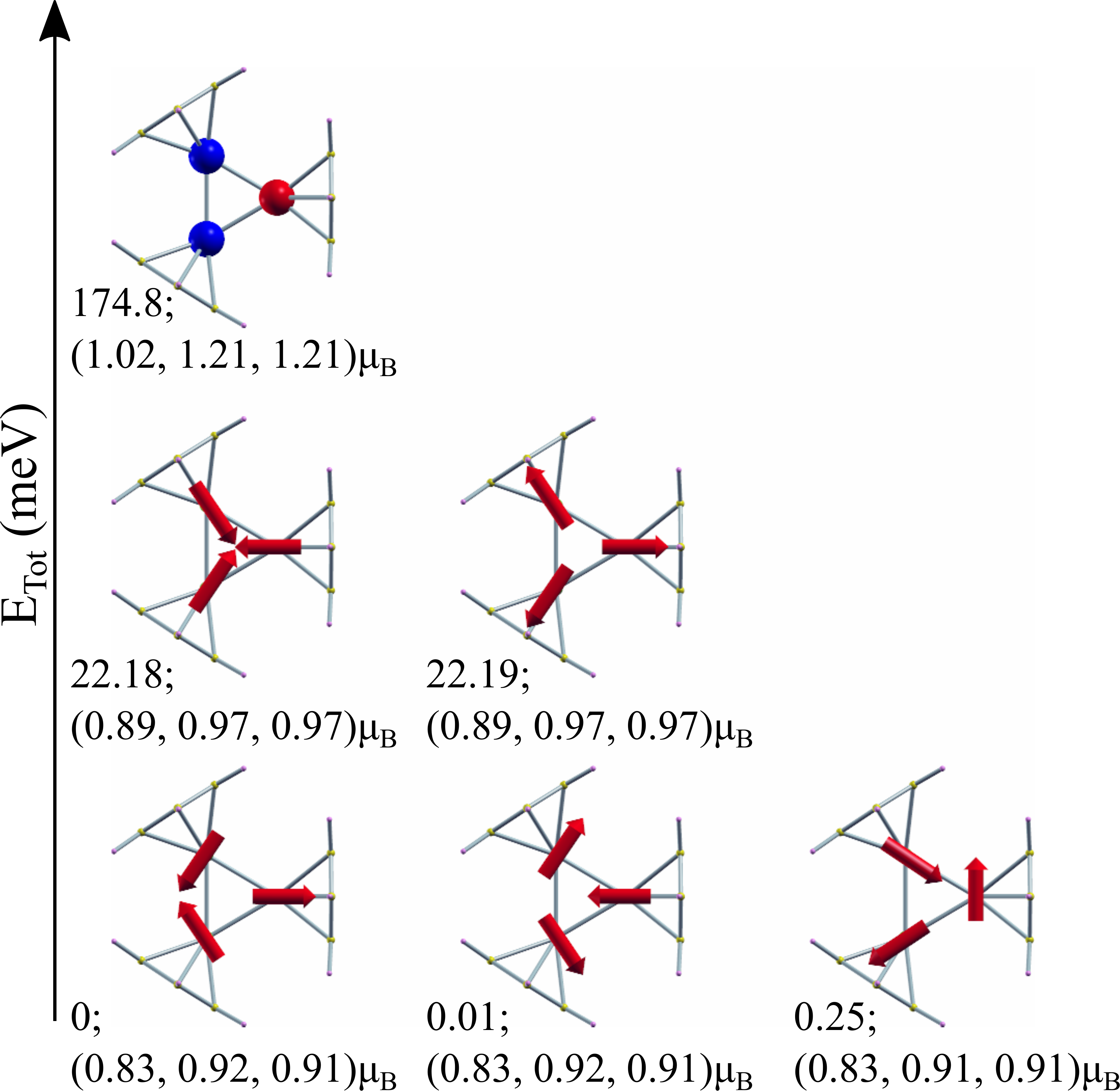} 
\caption{(Color online) Low energy magnetic configurations  of Co$_3$Bz$_3$. The red arrows point the directions of the local magnetic moments of the cobalt  atoms. Each label shows the energy difference (in meV) between the configuration shown and the ground state. In parentheses, the magnitude of the local magnetic moment at each cobalt atom (in units of $\mu_B$) is given in the atom order shown in Fig. \ref{fig1:esq}. Note that the non-collinear configurations become the ground state.
}
\label{fig2:scale}
\end{figure}

\subsection{\label{sec:relaxation} Collinear calculations}
We first consider the Co$_3$Bz$_3$ cluster with collinear magnetic moments.
We optimized the structures using SIESTA, then performed an exhaustive search to determine the ground state by exploring the different relative orientations of the benzene rings.
The geometric relaxations lead to the structure shown in Fig. \ref{fig1:esq}.  The ground state was found to be antiferromagnetic (AFM) with a total magnetic moment of $1\, \mu_B$, each cobalt atom contributing $\pm 1\,\mu_B$. The ground state corresponds to a structure with the cobalt atoms arranged in a slightly distorted equilateral triangle, with two bonds lengths of d$_{1,2}$ = d$_{1,3}$ = $2.34$ \AA~ and one of d$_{2,3}$ = $2.38$ \AA. The cobalt atoms  were situated in hollow positions with respect to the closest benzene ring,  $1.61$ \AA~ above  the hexagon barycentre.  
There were two higher energy ferromagnetic (FM) states, with total moments of 5 $\mu_B$ and 3 $\mu_B$, at $0.189$ and $0.286$ eV, respectively. 
These excited ferromagnetic configurations have been identified as the as the ground state in previous calculations.\cite{zhang2008structural}  
Note that when repeating the collinear calculations with  {\sc Quantum ESPRESSO} and VASP, we found similar AFM configurations as ground state.

The calculated Co$_3$Bz$_3$ ground state band gap was $1.26$ eV, and probably even larger because it is well known that band gaps calculated using the generalized gradient approximation are greatly underestimated.\cite{perdew1983physical,perdew1985density} The large band gap indicates that Co$_3$Bz$_3$ is very stable, in good agreement with the experimental results.\cite{kurikawa1999electronic}

We compared the results described above with the results of bare Co$_3$ cluster. We found that the bare Co$_3$ cluster formed a ferromagnetic isosceles triangle  with two bond lengths of $2.32$ \AA~ and one of $2.14$ \AA.
This can be ascribed to the Jahn-Teller effect distorting the bare Co$_3$ cluster.\cite{Gatteschi_Sessoli_book_2006} 
 The asymmetry in this structure was found to be reflected in the local magnetic moments, 
$2.61\,\mu_B$, $2.21\,\mu_B$, and $2.21\,\mu_B$, giving a total magnetic moment of $7$ $\mu_B$. This FM configuration was clearly preferred over the AFM configuration with an energy difference of $0.6$ eV.\cite{johll2011graphene}
These results were in clear contrast with the results for Co$_3$Bz$_3$. Adding  benzene molecules appears to almost remove the distortion of the Co$_3$ cluster. 
This could be partly due to charge transfer, because cobalt atoms gain about half an electron from the benzene molecules. To verify this hypothesis, we performed calculations for the charged Co$_3^{-0.5}$ cluster ground state and found that structural relaxation causes the cluster to tends toward an equilateral geometry.

\subsection{\label{sec:nocoll}Non-collinear calculations with spin-orbit interaction}
The antiferromagnetic ground state identified in the collinear calculations described in the previous section 
is certainly peculiar to Co$_3$Bz$_3$. The Co$_3$ cluster in particular and the nanostructures based in pure cobalt in general are ferromagnetic. Antiferromagnetic interactions and non-collinear magnetic landscapes may naturally give rise  to geometrical frustration as it is found in spin glasses.\cite{toulouse1987theory}
With these ideas in mind, we performed DFT calculations using a non-collinear full-potential linearized augmented plane-wave method for the previously described fully relaxed Co$_3$Bz$_3$ cluster.
 
We found an extra set of non-collinear states with in-plane magnetization, with the new ground state $174.8$ meV lower in energy  that the collinear AFM configuration.
The magnetic configurations could be divided into three groups according to the total energies and the orientation of the magnetizations of the cobalt atoms, as shown in Fig. \ref{fig2:scale}.
The first group contains low energy states with in-plane magnetic moments, aligned perpendicular or parallel to the benzene rings. 
The second group, with energies approximately $22$ meV higher than the ground state energy, has radial magnetization pointing either towards or away from the centroid of the cobalt atoms triangle. In both groups the magnetic moment per Co atom is between $0.83\,\mu_B$ and $0.97\,\mu_B$.
A third group is found with collinear states with magnetizations perpendicular to the cobalt atoms plane. The AFM configuration 
described in the collinear section 
lies $ 174.8 $ meV above the ground state, and the first FM state with a  magnetic moment of $5\,\mu_B$ was found at $948.5$ meV above the ground state.
It is noteworthy that in these cases, it was necessary to fix the direction of magnetization. Starting with a collinear configuration without fixing the directions of the magnetic moments led to the electronic structure calculations converging on non-collinear arrangements because of strong spin-orbit coupling of cobalt atoms.\cite{PhysRevLett.101.066402}

\begin{figure}[ht!]
\includegraphics[clip,width=0.5\textwidth,angle=0,clip]{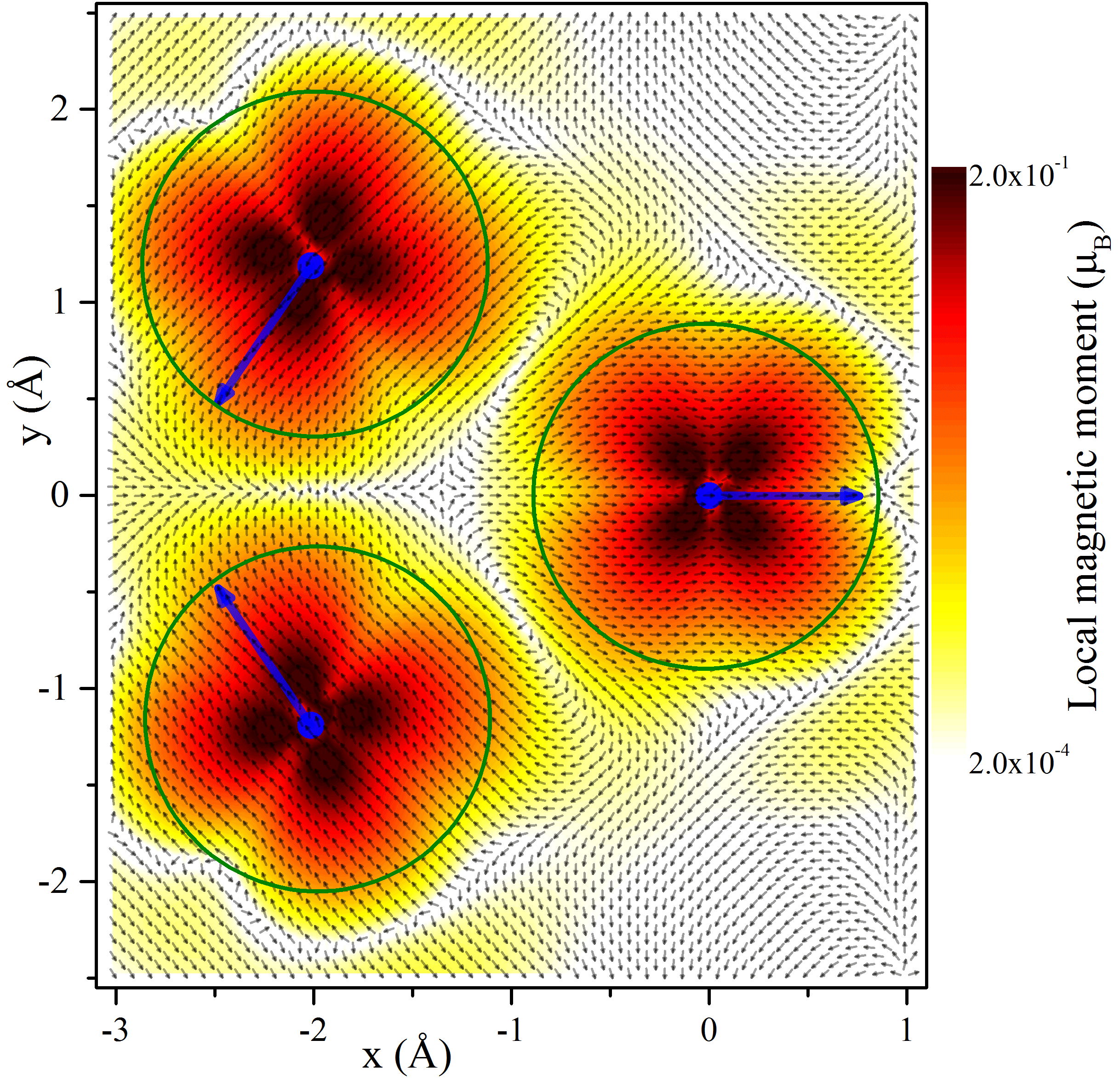} 
\caption{(Color online) Vector field  of  the  magnetization  density of the ground state in the cobalt atom plane for the Co$_3$Bz$_3$ ground state. The small arrows point in the direction of the local magnetic moment, and the colors indicate the magnitudes of the vectors. Each green circle indicates the muffin-tin radius of a cobalt atom, and the large arrow in each green circle indicates the total magnetic moment assigned to the cobalt atom. }
\label{fig3:map}
\end{figure}

The ground state presents an average orbital-magnetic moment $m_L = 0.06\,\mu_B$ and an average spin-magnetic moment $m_S = 0.07\,\mu_B$ per cobalt atom. 
The magnetization density in the cobalt plane is shown in Fig. \ref{fig3:map}.  
The magnetic moments are highly localized around the cobalt atoms. 
This strongly supports the idea of local moments used in the Heisenberg-like model described in the following section. 
In addition, the direction of the magnetization density rotates in the plane of the cobalt atoms generating a 
vortex-like pattern. Similar observations have been made for triangular and Kagome transition-metal 
networks\cite{sticht1989non,hobbs2000fully,hobbs2000fullyb} and in artificial Cr$_3$ structures.\cite{peralta2007noncollinear}

We tested the effects of stronger interactions between the 3d electrons in the cobalt atoms by adding the 
Hubbard $U$ and $J$ terms. We used typical values of $U=3$ eV for on-site Coulomb repulsion and $J=0.9$ eV 
for the exchange.\cite{piotrowski2011role}
The magnetic orders are qualitatively the same as the results shown in Fig.~\ref{fig2:scale} and they still are separable into the same three groups.
In particular, the ferromagnetic configuration is now at $799.4$ meV, while the antiferromagnetic is at $332.5$ meV above the ground state.
As a result of the Hubbard interaction, the localization of the d-states increased
with local magnetic moments slightly enhanced.

\section{\label{sec:modelH} Anisotropic Heisenberg Spin Model}
%
A  detailed description of the magnetization dynamics and the temperature dependence requires to take into account 
excited states.  Therefore, in this section we introduce a minimum quantum spin model capable of describing such magnetic properties.
The proposed Hamiltonian has the form
\begin{eqnarray}
H=J\left(\vec S_1\cdot \vec S_2+\vec S_1\cdot \vec S_3\right)+J'\vec S_2\cdot \vec S_3+\sum_{i=1}^3
 H_i,
\label{Hheis}
\end{eqnarray}
where $J$ and $J'$ are exchange constants, $\vec S_i$ is the spin operator corresponding to the 
$i$-cobalt atom (Fig.~\ref{fig1:esq}), and $H_i$ is the local magnetic anisotropy of this atom 
induced by the benzene rings. The Hamiltonian shown in Eq. (\ref{Hheis}) implies two basic assumptions. 
The first is that the spin of each cobalt atom is quantized and the magnitude of $S_i^2$ is conserved. 
This condition was fully satisfied for the low energy configurations shown in Fig. \ref{fig2:scale}.
The second assumption is that the magnetic moments are localized on the cobalt atoms.
This was supported by the magnetic solutions provided by the DFT calculations, as shown in Fig. \ref{fig3:map}. 
Moreover, the presence of two different exchange couplings $J$ and $J'$ is compatible with the 
\textit{isosceles}-type interactions indicated by the electronic structure 
calculations.\footnote{$J/J' = 1 $ indicates the limit of \textit{equilateral} interactions.}

The low energy configurations with non-radial symmetry (Fig.~\ref{fig2:scale}) indicate 
$S_i > 1/2$.\cite{Gatteschi_Sessoli_book_2006}
Here we take a spin $S_i=3/2$ for each cobalt atom.\footnote{In the case presented here, the main conclusions are 
not substantially different for a different spin of $S_i > 1/2$. For comparison, we also used a spin $S_i =2$,  without finding 
significant qualitative differences. 
A more detailed discussion of the spin value is given in the  Supplemental Material.} 
For the local magnetic anisotropy we use the lowest order terms compatible with the $C_{6v}$ symmetry,
$H_i\approx D\left(S_i^ {z_i}\right)^ 2+\dots$,\cite{Gatteschi_Sessoli_book_2006}
where $D$ is the uniaxial anisotropy parameter along the (local) high symmetry axis $z_i$.
These symmetry arguments are in good agreement with the DFT results for the CoBz cluster.
The total Hamiltonian, which needs all the spin operators in Eq. (\ref{Hheis}) to be written in a common reference frame,  
was solved through the numerical diagonalization of the $64\times 64$ Hamiltonian matrix.
We shall denote the eigenvalues and eigenvectors of $H$ by $\epsilon_N$ and $|N\rangle$ respectively.

The phase space of the model given in Eq. (\ref{Hheis}) was quite rich, as described in more detail in the Supplemental Material. 
We therefore searched for the parameters $J$, $J'$ and $D$ that better reproduced the main features of the DFT total energies. 
These features were in order of increasing energy, as shown in Fig~\ref{fig2:scale}:
(i) a ground state with magnetization in the cobalt plane,
(ii) a first excitation at very low energies, $\Delta_1= 0.2$ meV, with in-plane magnetization, 
(iii) a second excitation at $\Delta_2 = 22.2$ meV, 
(iv) and a first out-of-plane magnetization excitation $\Delta_3=174$ meV with antiferromagnetic order.

We found two possible scenarios with asymmetric interactions: $J/J' \approx 0.5$  (weak/strong/strong Co-Co interaction), and with  
$J/J'\approx 2$ (weak/weak/strong Co-Co interactions).
The best overall agreement according to the points mentioned above is found for the asymmetric configuration with $J/J'=2$ and $J\approx 5|D|$. 
Nevertheless, a configuration with $J/J'\approx 0.47$ and $J\approx |D|$ can not be fully discarded, so both are 
expected to be experimentally relevant under certain conditions.
This kind of duality has also been found experimentally in magnetic clusters containing triangular Mn$_3$ clusters.\cite{nguyen2016supramolecular}

Spin Hamiltonians are frequently used to connect experimentally observable quantities with the microscopic variables.\cite{Abragam_Bleaney_book_1970} 
For instance, a spin Hamiltonian  similar to Eq. (\ref{Hheis}) was  fitted to experimental data for a supramolecular aggregate containing SMMs, 
finding that it successfully explained the observed complex magnetic susceptibility and relaxation times behaviors.\cite{nguyen2016supramolecular}  
Accurate spectral information on the magnetic excitation of the clusters can be extracted using various techniques, including 
inelastic neutron scattering, optical spectroscopic methods, magnetic susceptibility measurements and electron 
paramagnetic resonance (EPR).\cite{Furrer_Waldmann_rmp_2013} 
Due to the energy scales involved in the problem, few of these techniques could provide meaningful magnetic information. 
The analysis of these characteristic scales lead us to consider EPR (see details in the Supplemental material).

By using perturbation theory in the \textit{ac} field, one can arrive to the following expression for the amplitude of the EPR signal
\begin{eqnarray}
{\rm Signal}\propto \sum_{N,M}\left|\vec B_{ac}\cdot \vec S_{MN}\right|^ 2(P_N-P_M)\omega^ 2f(\omega,\omega_{NM}),
\nonumber
\end{eqnarray}
where $P_M$ denotes the thermal equilibrium occupation of the $M$-energy level of Eq. (\ref{Hheis}), and $\vec S_{MN}=\langle M|\vec S_T|N\rangle$ with $\vec S_T$ the total spin. All the frequency dependence was included in the factor $\omega^2f(\omega,\omega_{NM})$.
A proper theoretical description of the EPR line-shape requires to account for the relaxation and decoherence mechanisms, 
which is beyond the scope of the present work. Instead, we used the frequently found experimental line-shape of 
$f(\omega,\omega_{NM})$, which can be approximated by the sum of two lorentzians centred at $\pm \omega_{NM}$ and 
with half-width at half-maximum of $T_2^{-1}$, the transversal relaxation rate.\cite{Abragam_Bleaney_book_1970} 
For a small frequency window the resonant states have a non-negligible contribution.
These states have   Bohr transition frequencies satisfying $|\omega-\omega_{N'M'}|T_2\ll 1$.
For low enough temperature only the ground state $N=0$ is occupied. The EPR signal is thus significant for the states with $M$ 
such that $|\omega-\omega_{0M}|T_2\ll 1$.\footnote{For our optimal parameters, at temperatures above $7$ K, the first 
excited state can have a significant population at $B_{dc}=10.2$ T. Thus, there may be a transition between 
the first and second excited states with a frequency around $138$ GHz. In the case of diluted clusters, 
one might expect $T_2\sim 1\;\mu{\rm s}-10$ ms, so resonant peaks will not overlap.}

\begin{figure}[t]
  \begin{center}
       \includegraphics[angle=0,clip,width=0.5\textwidth]{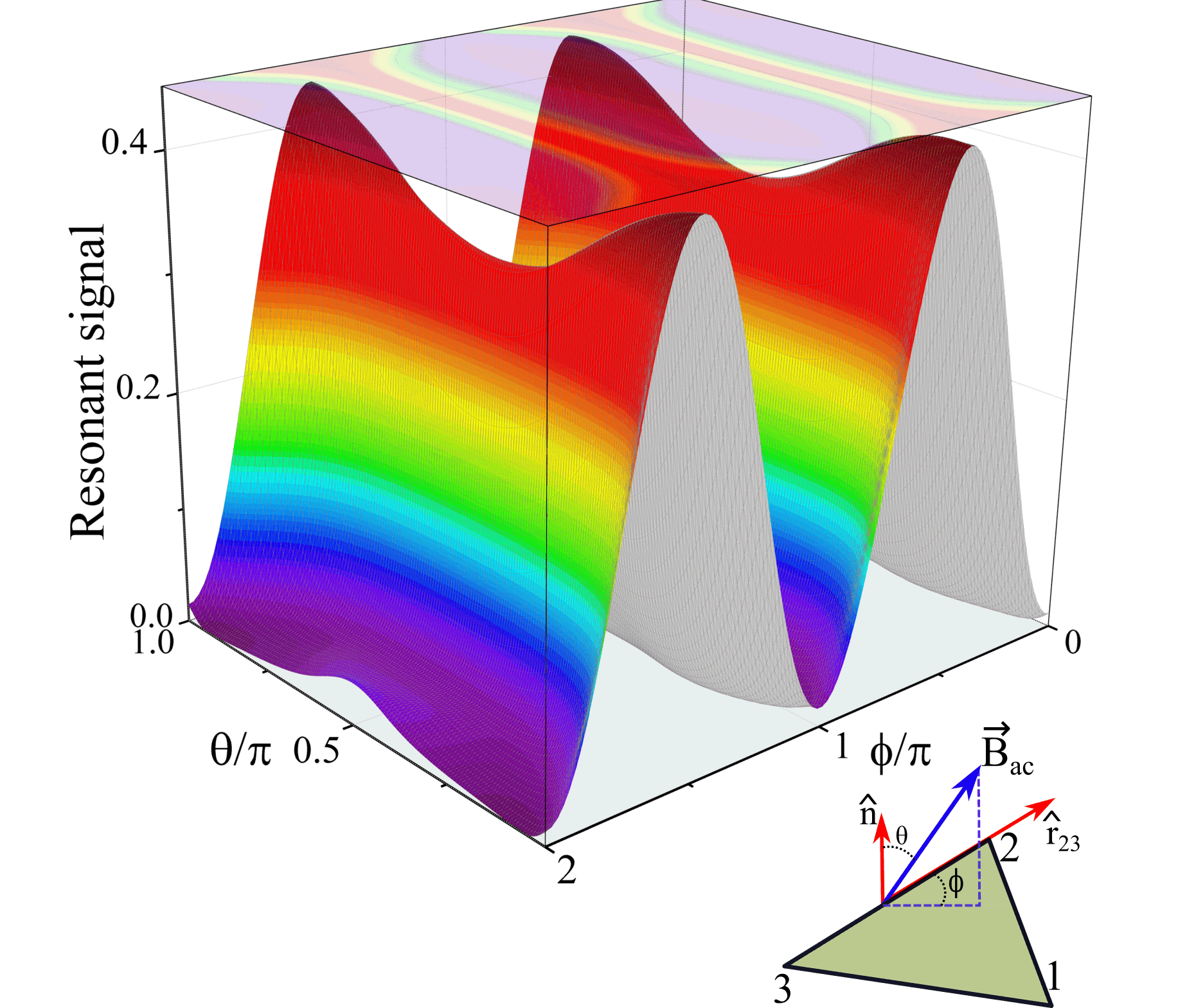}
  \end{center}
  \caption{(Color online) Contour plot of the on-resonant EPR signal, $\left|\hat B_{ac}\cdot \vec S_{0,1}\right|^ 2(P_0-P_1)$, versus the polar  $\theta$ and azimuthal $\phi$ angles (see lower inset). The ac field direction is parallel to $\vec B_{dc}\times \hat n$. The Hamiltonian parameters were taken for the optimal agreement with the DFT results and the temperature was fixed to 1 K.
 }
\label{FigS4}
\end{figure}

The angular dependence of EPR signal should reflect the magnetic anisotropy of the cluster. 
Here we focused on the transition between the ground state and the first excited state, although the excitation to the second excited state may be also probed.
We assumed a static \textit{dc} field  applied forming a polar angle $\theta$ with the normal $\hat n$ to the cobalt plane, an azimuthal angle $\phi$ with the $\vec r_{2,3}$ vector, see inset in Fig. \ref{FigS4}, and a small \textit{ac} field perpendicular to $\vec B_{dc}$.  
Figure~\ref{FigS4} shows the  angular dependence of the {\em on-resonant signal} $|\vec B_{ac}\cdot \vec S_{0,1}|^ 2(P_0-P_1)$. 
It is noteworthy that the angular dependence enters in the resonant signal on two counts. First, due to the magnetic anisotropy of the cluster, the excitation energy changes with the angle as reflected in the occupation difference $P_0-P_1$. 
This variation was observed in the signal dependence with temperature (Fig. S3 
of Supplemental Material).\footnote{The temperature dependence of the equilibrium 
occupations is the dominant one only for $T \lesssim 1-10$ K since above the broadening $T_2^{-1}$ 
can display a significant temperature variation.} 
Second, the oscillator strength of the transition changes through a variation of 
the $\vec B_{ac}\cdot \vec S_{0,1}$ matrix elements. We found that the resonant signal shows 
maxima when the \textit{dc} field is applied perpendicular to the cobalt atoms plane ($\theta=0,\pi$), 
and an \textit{ac} field pointing along  the easy axis of atom 1 
through coupling $J$ in Eq. (\ref{Hheis}). The resonant signal is more dramatically modulated in the $\phi$-angle. 
These angular dependences should be observable in EPR experiments.

\section{Final Remarks}
We investigated the prototype Co$_3$Bz$_3$ molecular cluster using several DFT methods (the SIESTA, VASP, {\sc Quantum ESPRESSO}, and ELK methods. We found that the benzene groups induce non-collinear magnetic configurations in the cobalt cluster. We demonstrated that the ground state of the cluster corresponds to configurations in which the magnetization remains in the plane defined by the three cobalt atoms.  This is especially striking when compared with the results for a bare Co$_3$ cluster, which has a ferromagnetic ground state with out-of-plane magnetization. We have therefore shown that the benzene rings play a crucial role in the magnetic landscape of the Co$_3$Bz$_3$ cluster.

To offer further insights into the responses of the magnetic properties, we built a quantum spin model Hamiltonian that accounted for the low energy magnetic solutions of the electronic structure calculations. This Hamiltonian contains local uniaxial magnetic anisotropy, associated with the presence of the benzene rings, and antiferromagnetic coupling between the spins of the Co atoms. This approach leads to two possible stable configurations 
corresponding to different deviations from perfectly symmetric interactions.
In contrast to the symmetric case, which displays spin frustration in the Heisenberg limit,\cite{Hsieh_Rene_prb_2012} the anisotropic and asymmetric molecule leads to an effective total spin of $S=3/2$. 

A detailed study of the energy scales lead us to propose EPR as the most suitable technique to analyze the spectral magnetic features of the Co$_3$Bz$_3$ cluster within the reach of experimentally realizable conditions.
In addition, we demonstrated that the Co$_3$Bz$_3$ cluster has a preferential in plane anisotropy, revealed through a drastic azimuthal angular dependence between the direction of the radiofrequency field and the easy axis of each Co atom.
A meticulous experimental determination of the magnetic anisotropy of the molecule would be even possible, as done for instance in single molecule tunnel junctions based on organic molecules.\cite{Gaudenzi_Burzuri_nanolett_2016,Frisenda_Gaudenzi_nanolett_2015}

During the preparation of this manuscript S.T. Akin {\em et al.}\cite{akin2016size} reported XMCD measurements for Co$_3$Bz$_3^+$ cluster with null spin-magnetic moment. The null signal could be understood as a consequence of the predicted 
non-collinear character of the local magnetic moments. In order to elucidate the reason behind the absent of net magnetic moment, further experiments are required. The magnetic response of the EPR could distinguish between the collinear and non-collinear scenarios.

\begin{acknowledgments}
This work has been partially supported by the Projects FIS2013-48286-C02-01-P and FIS2016-76617-P of the Spanish Ministry of Economy and Competitiveness MINECO, the Basque Government under the ELKARTEK project(SUPER), and the University of the Basque Country  (Grant No. IT-756-13). TA-L acknowledge the grant of the MPC Material Physics Center - San Sebasti\'an. FA-G acknowledge the DIPC for their generous hospitality. Authors also acknowledge the DIPC computer center and the useful discussions with L. Chico.
\end{acknowledgments}



\pagebreak

\begin{center}
\textbf{\large Supplemental Materials: Non-Collinearity in Small Cobalt-Benzene molecular Clusters}
\end{center}
\setcounter{figure}{0} 
\setcounter{section}{0} 
\setcounter{equation}{0}
\setcounter{page}{1}
\renewcommand{\thepage}{S\arabic{page}} 
\renewcommand{\thesection}{S\Roman{section}}   
\renewcommand{\thetable}{S\arabic{table}}  
\renewcommand{\thefigure}{S\arabic{figure}} 
\renewcommand{\theequation}{S\arabic{equation}} 

\section{\label{SIESTA_detalle}DFT Details}

\subsection{LCAO methods: SIESTA}
SIESTA calculations were performed using the exchange and correlation potentials with the generalized gradient approximation (GGA) in   Perdew-Burke-Ernzenhof form.\cite{perdew1996generalized}
We used an electronic temperature of $25$ meV and a mesh cutoff of $250$ Ry for all the calculations. The atomic cores were described using nonlocal norm-conserving relativistic Troullier-Martins\cite{troullier1991efficient} pseudopotentials with non-linear core corrections factorized in the Kleynman-Bylander form.
%
The pseudopotentials were tested to ensure that they accurately reproduce the eigenvalues of different excited states of a bare atom. We tested the cobalt pseudopotential  against the bulk cobalt. We use the 4s$^{1}$3d$^{8}$ valence configuration, and found that the ground state was the hcp structure, which had a stability of $0.025$ eV/atom greater than that of the fcc structure. For the hcp structure, the first neighbor distance was $2.54$ \AA~ and the magnetic moment per atom was $1.64$ $\mu_{B}$, in good agreement with experimental results.\cite{myers1951spontaneous} The cobalt pseudopotential was described in detail and used to study Co$_{13}$ clusters adsorbed onto graphene in a previous study.\cite{alonso2015chemical}
 The basis sets for cobalt, hydrogen, and carbon were  the double-polarized basis sets identified in previous studies. 

Geometric relaxations used  conjugate gradient structure optimization. 
The hydrogen atoms of the benzene rings made it necessary to impose small convergence thresholds on forces of the order of $10^{-3}$ Ry/bohr $\approx 2.57\times 10^{-3}$ eV/\AA. We found that imposing less strict convergence conditions over both the energies and the forces may erroneously lead to deformed structures.
In each case, a single $\Gamma$-point was chosen for the calculations and  $15$ \AA~  of empty space was added to avoid interactions between nearest-neighbors cells. 

\subsection{Plane-waves methods: Quantum ESPRESSO and  VASP}
Plane-wave methods such as {\sc Quantum ESPRESSO}\cite{QE-2009} and VASP\cite{VASP,VASP2} were used. Self-consistent calculations were performed very accurately using a plane-wave  kinetic energy cutoff  of $80$  Ry. As in the SIESTA calculations, the generalized gradient approximation in exchange-correlation was in Perdew-Burke-Ernzerhof (PBE)\cite{perdew1996generalized} form. Norm-conserving Troullier-Martins pseudopotentials\cite{troullier1991efficient} were used in Quantum ESPRESSO, and a projector augmented wave potential construction was used in VASP.
All the atoms were allowed to relax by conjugate gradients until the forces converged, with a tolerance of $10^{-4}$  eV/\AA. We then checked that the relaxed geometries reproduced the SIESTA results.

\subsection{Full potential methods: ELK}
An all-electron full-potential linearised augmented-plane wave (FP-LAPW) were performed using the ELK code. The electronic exchange-correlation potential was treated within the local spin density approximation (LSDA) \cite{von1972local} to avoid gradients effects of the generalized gradient approximation in the non-collinear calculations.
Wavefunctions in the interstitial density and potential were expanded into plane waves with a wavevector cutoff of  $k_{max} =15/R_{Co}$, where $R_{Co}\approx 1.16$ \AA~ is the muffin-tin radius of cobalt.

\section{Spin Hamiltonian}

\subsection{Local spins and symmetry arguments}

The DFT results indicated that the energy levels corresponding to the $d$-orbitals of the cobalt atoms in a CoBz 
cluster split in two, with the lowest orbital triplet almost fully occupied and a half-occupied excited doublet. 
Thus, the filling of the levels following the Hund's rule leads to a spin $S=3/2$. 
Although the degeneracy of the lowest orbital 
triplet is partially broken due to the new crystal field in the Co$_3$Bz$_3$ cluster, the qualitative filling is 
essentially conserved.  Consequently, we assumed later that each cobalt behaves as a $S=3/2$ spin. 
This spin is also compatible with the $^4$F ground state observed in the gas phase.\cite{NIST_ASD}

In order to include the local magnetic anisotropy, we made the following symmetry considerations.
An isolated CoBz unit has C$_{6v}$ symmetry. The local spin Hamiltonian compatible with this symmetry can be written in terms of the (tesseral tensor) Stevens operators $\hat O_k^q\left(S\right)$ as\cite{Abragam_Bleaney_book_1970}
\begin{eqnarray}
H_i =\sum_{k=2,4,6}\sum_{q=-k}^{k} B_k^q\hat O_k^q\left(S_i\right),
\label{Hani}
\end{eqnarray}
where $B_k^ q$ are real coefficients. $\hat O_k^q\left(S\right)$ are in turn linear combinations of spin operator components. The lowest order non-isotropic term corresponds to the operator $\hat O_2^0 \propto  (S_i^{z_i})^2$, which is the uniaxial term used in this work. 
Higher (even)  powers of  $S_i^ {z_i}$  do not introduce any 
qualitative change to the energy spectrum and  they can be thought as a renormalized $D$-value, hence we  neglected them. All these terms commute with $S_i^ {z_i}$,  and thus, they do not introduce mixing between the eigenstates of $S_i^{z_i}$.  Importantly, the lowest transverse terms compatible with the C$_{6v}$ symmetry of CoBz 
involves the sixth power of the ladder operator, $\left(S_i^ {\pm}\right)^ 6$. They therefore do not contribute to the spectrum of a spin $S\le 5/2$, so we discarded them.

\subsection{Spin Hamiltonian parameters}
As stated in the main text,  the spin Hamiltonian Eq. (1) 
contains three fitting parameters, $J$, $J'$ and $D$, for a given local 
spin $S$. Herein, we  described in detail the derivation of the optimal 
parameters that best match the information provided by the DFT calculations.

The electronic structure calculations of a single CoBz molecule gave an 
energy difference of $8.2$ meV between configurations with the magnetization
out of plane (lowest energy) and  in-plane, suggesting that the easy axis would 
be out of plane with $D\approx -4.1$ meV for $S=3/2$. The DFT results for the CoBz complex 
show that changing the magnetization orientation along the benzene plane did 
not significantly change the energy of the complex, in good agreement with 
our symmetry arguments.
We found that the benzene rings are not deformed appreciably when a Co$_3$Bz$_3$ 
cluster is formed, so it may be tempting to use the local anisotropy of each CoBz unit.
 However, this option was strongly discouraged by the DFT results. First, the 
distances between the cobalt atoms (2.34 \AA) were of the same order than the distance 
between a cobalt atom and the closest carbon, approximately $ 2$ \AA.  Second,
the charge of the cobalt ions in a CoBz and the Co$_3$Bz$_3$ cluster differ 
by half an electron. 
Third, the distances between cobalt atoms and the closest benzene plane changes by as much as 9$\%$.

The most important characteristics  to reproduce from the DFT results, shown in
Fig.~(2) 
\begin{itemize}
\item  a ground state with magnetization in the cobalt atom plane,
\item a first excitation at  $\Delta_1  \sim 0.2$ meV, with in-plane magnetization,
\item a second excitation at $\Delta_2 \sim 22$ meV (in-plane magnetization), and
\item a third excitation at $\Delta_3 \sim 72$ meV with out-of-plane magnetization.
\end{itemize}

\begin{figure}[t]
  \begin{center}
    \includegraphics[angle=-90,clip,width=0.55\textwidth]{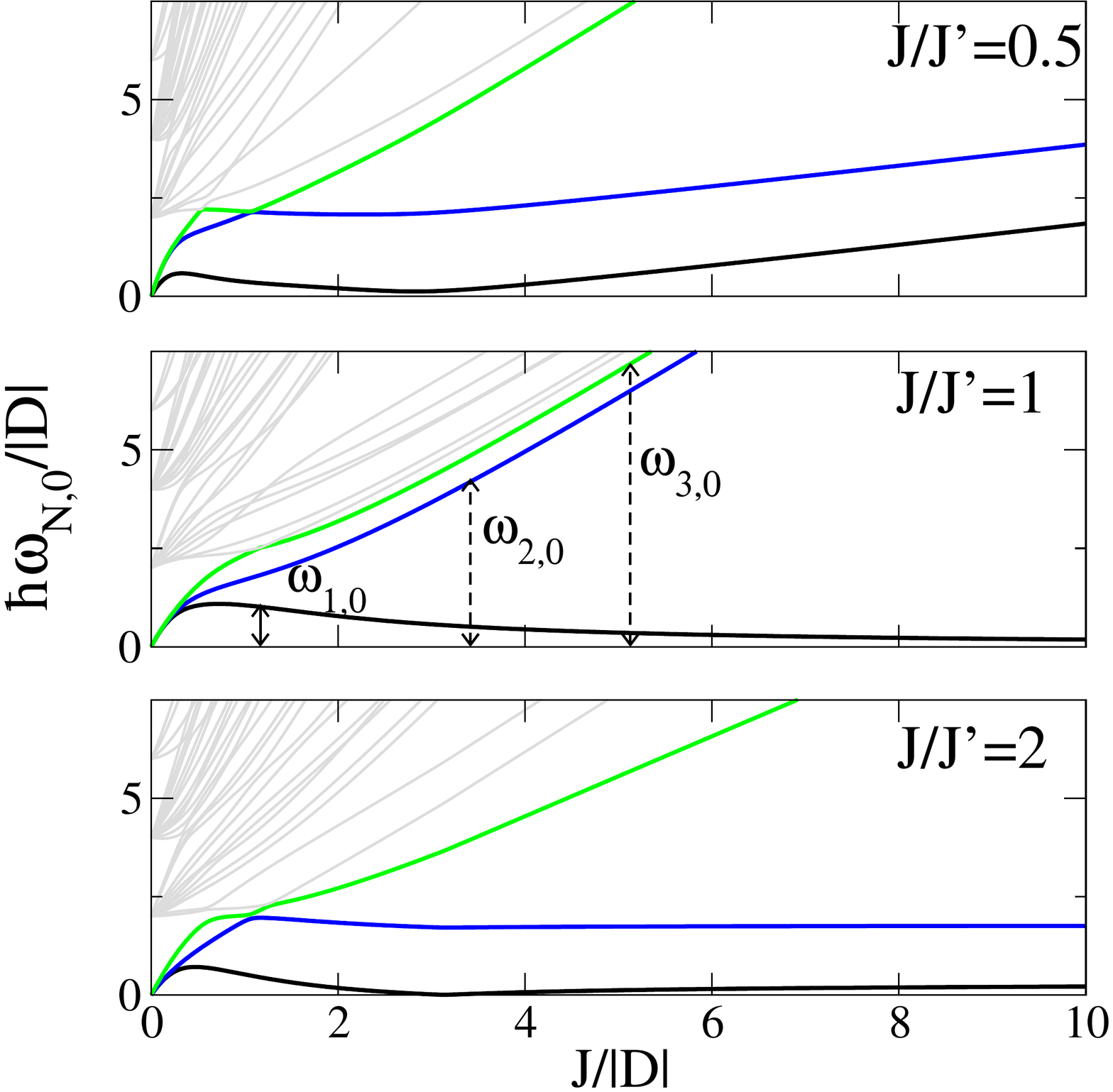}
  \end{center}
  \caption{(Color online) Excitation energies from the ground state vs.  $J/|D|$ for three different $J/J'$ ratios. 
  $\hbar\omega_{1,0}$ (black line) is identified with the 0.25 meV non-collinear excitation in Fig.~(2) 
   $\hbar\omega_{2,0}$ (blue line) with the 22 meV excitation, and $\hbar\omega_{3,0}$ (green line) 
   with the antiferromagnetic configuration at $72$ meV.   }
\label{figS1}
\end{figure}

 Our approach to set the three parameters consisted of the following steps.
First, for three different $J/J'$ scenarios,  we looked at the qualitative behavior of the 
energy spectra with the $J/|D|$ ratio between the limits of the isolated CoBz units ($J/|D|=0$), 
and the isotropic cluster ($J\gg |D|$).   Crucially, the magnetic configurations resulting 
from the non-collinear calculations indicated the dominance of spin-exchange interactions over 
anisotropy, so the most natural scenario corresponded to $J\gtrsim |D|$. The excitation  
energies $\hbar\omega_{N,0}$
 from the ground state are plotted against $J/|D|$ for three different  $J/J'$ values in Fig. \ref{figS1}. 
We associated the first excitation with the DFT electronic state at $0.2$ meV above the ground state.
 The second energy level was ascribed to the configurations found $22.2$ meV above the ground state. 
Thus, we choose the parameters such that $\omega_{1,0}/\omega_{2,0}=\Delta_1/\Delta_2\approx 0.01$
while keeping the gap $\hbar\omega_{2,0}$ as large as possible in terms of $D$ to avoid unrealistically l
arge anisotropy barriers.
The third magnetic configuration  at  $ 72$ meV, identified from the DFT results as an AFM configuration with out-of-plane magnetization, was taken as the third excited state of the model. Comparing the energy spectra shown in Fig.~(2) 
and  Fig. \ref{figS1}, we observe that the first two conditions are quantitatively satisfied for $J/|D|\approx 2.5$ in the asymmetric case with $J/J'\approx 0.5$ or
 for $J/|D|\in \left[2.5,\,6\right]$ with $J/J'=2$. The symmetric configuration $J=J'$ did not provide a qualitative agreement for any  $J/|D|$ ratio.

\begin{figure}[t]
  \begin{center}
    \includegraphics[angle=-90,clip,width=0.55\textwidth]{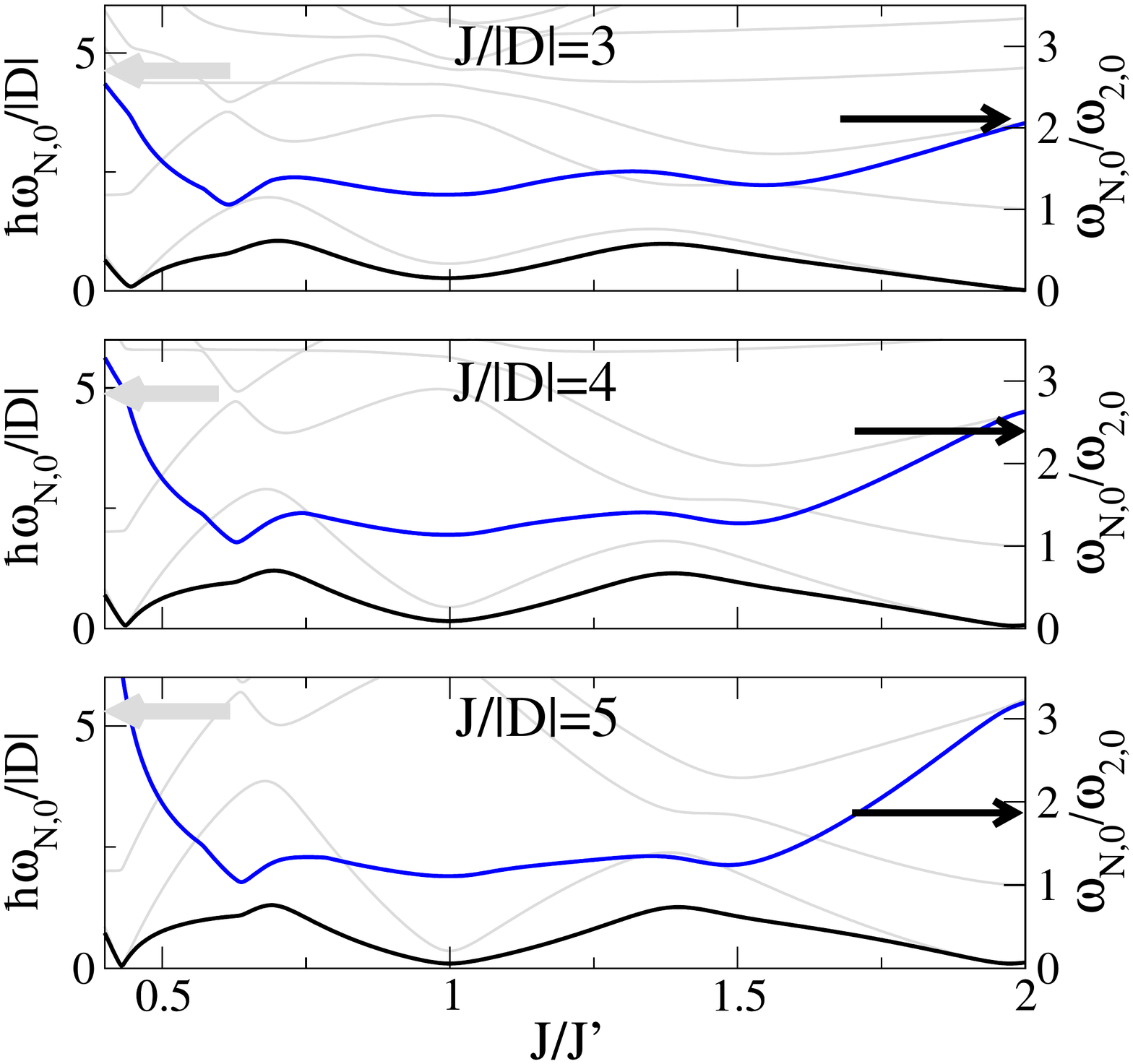}
  \end{center}
  \caption{ (Color online)  Excitation energies (gray lines, left axis) from the ground state vs. the ratio $J/J'$ for three different ratios $J/|D|$. 
The ratios  $\omega_{1,0}/\omega_{2,0}$  and
  $\omega_{3,0}/\omega_{2,0} $ (right axis) are plotted as thick blue and black lines respectively.
  These ratios approach the DFT results of Fig. 2 when $J/J'=2$ for $J/|D|=5$.} 
\label{figS2}
\end{figure}

A more quantitative picture is obtained by inspection of the excitation spectrum versus $J/J'$, shown in Fig.~\ref{figS2}. 
 The $\omega_{1,0}/\omega_{2,0}$ and $\omega_{3,0}/\omega_{2,0}$
 ratios were plotted versus $J/J'$ for three different values of $J/|D|$, with the ideal ratios close to 0 and 3.3, respectively. For reference, the corresponding excitation spectra are plotted in grey. The best agreement is found for $J/J'\approx 0.47$ and, especially, for $J/J'\approx 2$.  Again, the symmetric case ($J=J'$)  was far from the ideal ratios 
 for any $J/|D|$ in the considered range of parameters.

Summing up, optimal agreement with the DFT results was found for $(D,\,J,\,J')=(-12.64,\, 63.2,\, 31.6)$ meV and $(D,\,J,\,J')=(-12.64,\, 63.2,\, 134)$ meV.  These values are higher than those typically found within SMMs.  However, the cobalt centers in the Co$_3$Bz$_3$ cluster are much closer, so a direct exchange mechanism is possible. Thus, it seems clear that both local magnetic anisotropy due to the benzene rings and the direct-exchange between the local  moments associated to the Co atoms plays a crucial role.

\subsection{Connection to experiments}
Before trying to connect to an specific measurement, it is worth looking at the energy scales involved. If we denoted the excitation energies from the ground state to the first and second excited states as $\hbar\omega_{1,0}$ and $\hbar\omega_{2,0}$ respectively, and introducing the temperature $T_M=\hbar\omega_{M,0}/k_B$, magnetic field $B_M=\hbar\omega_{M,0}/(g\mu_B )$, and frequency $f_M=2\pi \omega_{M,0}$, we got the following energy scales at zero external field: $\hbar\omega_{1,0}\sim 0.2$ meV ($T_1\sim 2$ K, $B_1\sim 2$ T, and $f_1\sim 50$ GHz), and  $\hbar\omega_{2,0}\sim 22$ meV ($T_2\sim 250$ K, $B_2\sim 190$ T, and $f_2\sim 5$ THz). This had the following consequences for the possible experimental observation. 
Static measurements of the susceptibility may provide significant information about the lowest energy excitation for $T\lesssim 2$ K. Furthermore, for $T$ close to room temperature there may be other excitations not included here, like phonons. We had not found significant changes of the susceptibility with temperature or magnetic fields  below 10 T. 
\textit{ac}  measurements of the dynamical susceptibility are also commonly used to extract additional magnetic information of SMMs, but the frequency range is limited to $f\in [1\;{\rm Hz}-0.1\;{\rm MHz}]$,\cite{Repolles_Cornia_prb_2014} clearly outside the energy range of interest. High $dc$-field EPR measurements may provide an accurate spectral information. In fact, for the $B_{dc}=10.2$ T $J$-band, the electron spin resonance is found around $285$ GHz.\cite{grinberg2013very}  For such an applied external field applied out of the Co's plane, $\hbar\omega_{1,0}\sim 1.1$ meV and $\hbar\omega_{2,0}\sim 1.2$ meV ($275-286$ GHz), within the range of experimental frequencies.

\subsection{Analysis of the EPR signal}
EPR experiments apply a fixed \textit{dc} field along a given direction, which we defined as the quantization axis $z$, and a small perpendicular \textit{ac} field $\vec B_{ac}(t)=\vec B_{ac}\sin\omega t$, along the $x$-axis. For the Co-Bz molecule, the most convenient set-up corresponds to the $J$-band ($B_{dc}=10.2$ T) with frequencies $2\pi\omega\approx 285$ GHz. We used time dependent perturbation theory because the perpendicular \textit{ac} field is much smaller, with typical intensities of the order of 1 G. 

\begin{figure}[t]
  \begin{center}
       \includegraphics[angle=0,clip,width=0.5\textwidth]{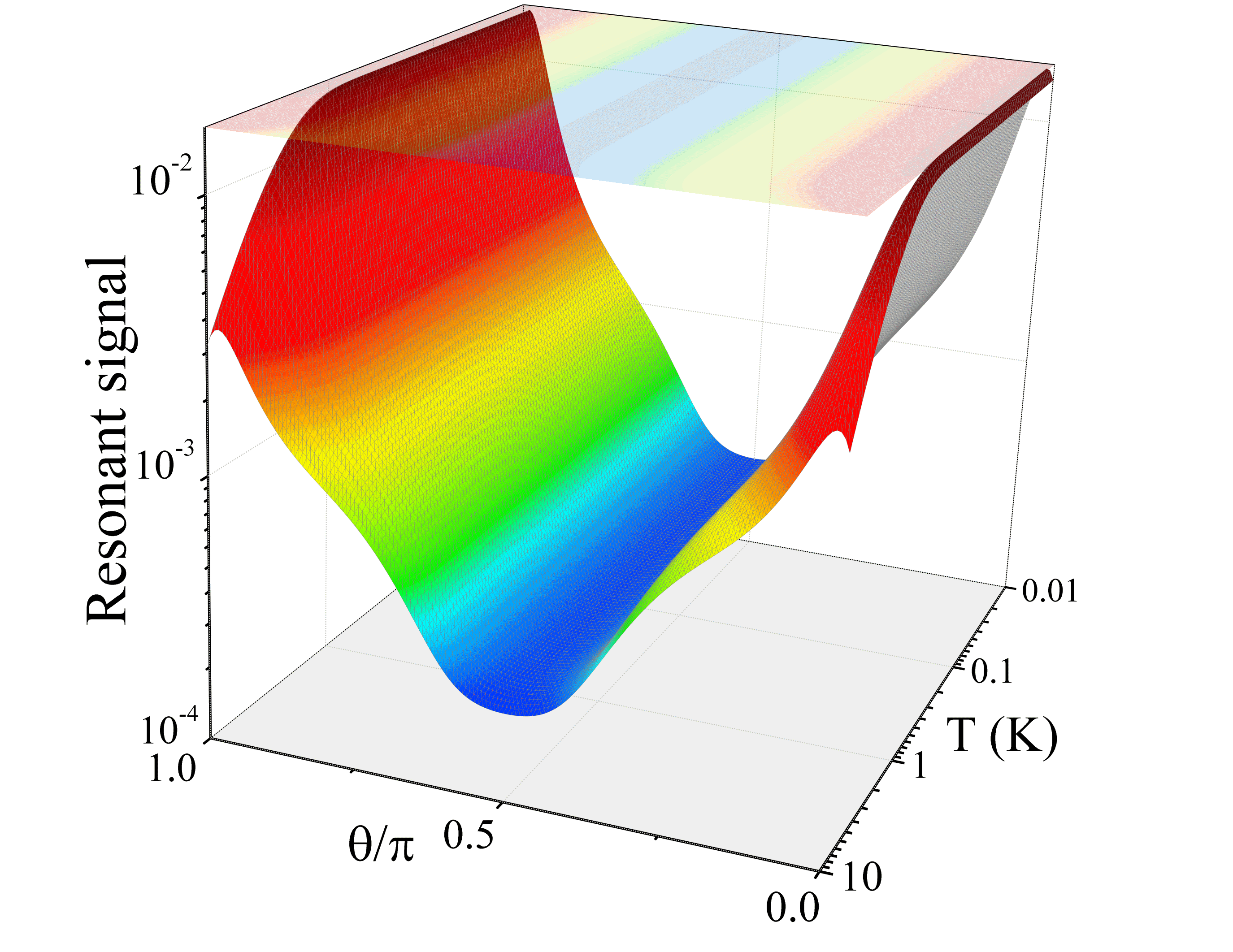}
  \end{center}
  \caption{(Color online) Contour plot of the on-resonant EPR signal, $|\hat B_{ac}\cdot \vec S_{1,0}|^ 2(P_1-P_0)$, versus the angle  $\theta$ formed by the normal to the Co's plane and the \textit{dc} magnetic field ($\hat B_{ac}=\vec B_{ac}/|\vec B_{ac}|$ ). The \textit{ac} field is parallel to one of the easy axis of atom 1 in Eq. (1)]. 
 }
\label{fig4M}
\end{figure}

The absorbed power is given as the variation of the average energy $\langle\langle H(t)\rangle\rangle={\rm Tr}[\hat\rho(t)\hat H(t)]$ with time, where $\hat\rho(t)$ is the density matrix.
We here assumed a coherent dynamics. We used an interaction picture with respect to the term $H'(t)=g\mu_b\vec B_{ac}\cdot \vec S_T\sin\omega t$ and a first order perturbative expansion for the density matrix operator. Thus, the instantaneous power was given by
\begin{eqnarray}
W(t)&\equiv&\frac{d\langle\langle H(t)\rangle\rangle}{dt} \approx
\omega \gamma B_{ac} \cos\omega t\sum_N P_N S^x_{N,N}
\crcr
&+&
\frac{\omega\gamma^2B_{ac}^2}{\hbar^2}\sum_{NM}\left|S^x_{NM}\right|^2(P_M-P_N)
\crcr
&&\hspace{-1.5cm}\times
\left[\cos t\omega\sin t\omega_{NM}+\frac{\omega_{NM}}{\omega}\left(
\cos t\omega_{NM}\sin t\omega
-\sin 2t\omega\right)\right],
\crcr
&&
\label{IPower}
\end{eqnarray}
where $S^x_{NM}=\sum_l \langle N|S^x_l|M\rangle$, $\gamma=g\mu_B$ is the gyromagnetic ratio, and $P_M$ denotes the thermal equilibrium occupation of the energy level $M$.

There are however two considerations to be made in order to connect the absorbed power $W(t)$ with the EPR signal. First, the absorption signal of the EPR measurement corresponds to the average of instantaneous power over a measurement time $\tau\gg 2\pi/\omega$.  Second, the coupling of the spin system with 
the environment induces a dissipative dynamics. When two energy levels are close to resonance for a given frequency $\omega$, i.e., $|\omega-\omega_{NM}|/\omega\ll 1$, the incoherent dynamics introduces two new time scales, the {\em longitudinal or relaxation time} $T_1$ and the {\em transversal or decoherence time} $T_2$.\cite{Abragam_Bleaney_book_1970} These two time scales determine the line-shape of the EPR absorption spectra and, in particular, the width of the resonances, given essentially by $T_2^ {-1}$.

When $W(t)$ is averaged over a long period of time $\tau\gg 1/\omega$,  the first term cancels. For a frequency window
$|\omega-\omega_{NM}|T_2\ll 1$,  the  {\em steady-state} absorbed power is then proportional to $\gamma^ 2 B_{ac}^ 2|S^x_{N,M}|^2(P_N-P_M)$. The incoherent dynamics modifies the time dependence in the last part of Eq. (\ref{IPower}) which, in the frequency domain, leads to a finite resonant amplitude and a finite frequency width. We approximated this frequency dependence using the collision broadened profile of Van Vleck and Weisskopf (1945),\cite{Abragam_Bleaney_book_1970}  
\begin{equation}
\bar W (\omega)=\frac{\gamma^2B_{ac}^2}{\hbar^2}\left|S^x_{NM}\right|^2(P_M-P_N)\omega^2f(\omega,\omega_{NM}),
\end{equation}
where $f(\omega,\omega_{NM})$ takes the form
\begin{eqnarray}
\frac{1}{\pi} \left[
\frac{T_2^{-1}}{(\omega-\omega_{NM})^2+T_2^{-2}}+
\frac{T_2^{-1}}{(\omega+\omega_{NM})^2+T_2^{-2} }
\right].
\nonumber
\end{eqnarray}
However, we should keep in mind that this EPR line shape does not account for saturation effects and the associated 
temperature increment, which for a typical \textit{ac} field of $0.1$ G occurs for longitudinal relaxation time $T_1\gtrsim 1\;\mu$s.


\end{document}